# Encoding of low-quality DNA profiles as genotype probability matrices for improved profile comparisons, relatedness evaluation and database searches

K. Ryan, D. Gareth Williams[a], David J. Balding[b,c]

a. DGW Software Consultants LTD

b. UCL Genetics Institute, University College, London WC1E 6BT

c. Centre for Systems Genomics, School of BioSciences and School of Mathematics and Statistics, University of Melbourne, VIC 3010, Australia

**ABSTRACT**

Many DNA profiles recovered from crime scene samples are of a quality that does not allow them to be searched against, nor entered into, databases. We propose a method for the comparison of profiles arising from two DNA samples, one or both of which can have multiple donors and be affected by low DNA template or degraded DNA.  We compute likelihood ratios to evaluate the hypothesis that the two samples have a common DNA donor, and hypotheses specifying the relatedness of two donors. Our method uses a probability distribution for the genotype of the donor of interest in each sample.  This distribution can be obtained from a statistical model, or we can exploit the ability of trained human experts to assess genotype probabilities, thus extracting much information that would be discarded by standard interpretation rules. Our method is compatible with established methods in simple settings, but is more widely applicable and can make better use of information than many current methods for the analysis of mixed-source, low-template DNA profiles. It can accommodate uncertainty arising from relatedness instead of or in addition to uncertainty arising from noisy genotyping. We describe a computer program GPMDNA , available under an open source license, to calculate LRs using the method presented in this paper.

## 1  Introduction

Crime scene samples often contain DNA from multiple individuals, some or all of whom contribute small amounts of DNA that may have suffered degradation following environmental exposure. An electropherogram (epg) arising from such a sample can show stochastic effects that make it difficult to identify all of the alleles in the underlying genotypes [1,2].  Uncertain alleles are often encoded as "unknown". For example the UK National DNA Database uses the coding F which can match any allele in a subsequent database search [3].  But if there is only limited uncertainty about the genotypes at several loci,  designating them all as unknown is wasteful of potentially valuable information. We propose instead to encode the epg information as probability distributions over genotypes, each encoded as a genotype probability matrix (GPM).

We show how to compute from two GPMs a likelihood ratio (LR) assessing the evidence that the donors of interest to the samples have a specified genetic relationship, including the important special case that they are the same person. In standard settings there is a reference profile that specifies a unique genotype, while the information from the crime scene epg may be represented as a list of alleles with peak heights.  We allow each of the two profiles being compared to be represented as a GPM.  In the special case of a reference profile, one genotype is assigned probability one.

We expect that our proposed method will generate substantial benefits from its use as an intelligence tool. Given one or more complex profiles, the method permits speedy and accurate inferences to guide an investigation, for example in seeking a contributor to multiple, noisy crime scene profiles (CSP). However, our approach is also suitable for conveying weight of evidence to a finder of fact within a judicial system. We believe that the approach has great potential in improved database searches: the CSP and partial profiles in the database can be represented as GPMs, leading to informative search results that take account of any uncertainty in either query or database profiles. Rigid rules for the "quality" of a profile to be included in the database, or for searches of it, can be greatly relaxed. With probabilistic encoding of epg information, poorer quality samples will automatically generate more diffuse probability distributions and hence weaker LRs, without the need for rules based on thresholds. Our approach can also be used when uncertainty arises due to genotype information only being available for relatives of the target individual, including familial database searches.

We briefly discuss below the generation of a GPM from an epg using a statistical model, but our approach can also be used in conjunction with an expert forensic scientist who directly encodes the epg information as a GPM. This encoding implies an element of subjectivity, but inter-expert comparisons and blind proficiency testing can satisfy a court's requirement for objectivity. Our approach is designed to exploit the ability of experts to make use of complex information quickly and efficiently, without the need for inflexible rules or sophisticated software. Experts can make better use of the epg than is the case in much current practice because probabilities allow gradations of judgement that better reflect the available information than do interpretation rules based on thresholds.

In Appendix A4 we describe GPMDNA, a software tool that implements the methods described in this paper.

## 2 Encoding an epg in a genotype probability matrix

Although the interpretation of an epg can largely be automated, an expert forensic scientist is usually called on to confirm the automatic interpretation, and sometimes over-rule decisions relating to apparent artefacts such as split peaks and stutter. We distinguish this process of "interpretation" from that of "evaluation" in which numerical summaries of evidential weight are computed. For a single-source sample of good quality DNA, the result of an interpretation is the genotype of the donor at the tested loci. However, for many CSPs the genotype of an unknown donor of interest cannot be inferred with certainty from the epg. Reasons for this can include low DNA template and/or degraded DNA leading to the possibility of dropout of some alleles, DNA from more than one individual (mixed profile), and uncertainty about possible artefacts.

The need for objectivity in legal settings has led to the adoption of interpretation rules based on thresholds that have been shown in laboratory trials to have good properties over many real or simulated examples [4, 5]. For example, peaks below a threshold (typically 25 to 50 relative fluorescence units, RFU) are dismissed as noise, while single peaks above another threshold (typically 200 to 300 RFU) are interpreted as homozygotes (thus ruling out dropout of another allele from the donor at that locus) [3]. Similarly, a peak is interpreted as stutter or allelic according to whether or not its height is below a threshold fraction of the peak height at the "parent" allele (at one repeat unit larger than the peak under consideration).

These interpretation rules suffer from a "cliff edge" effect [6]. For example, a peak in a stutter position can be classified as certainly stutter, yet if it were slightly higher it would be regarded as certainly allelic. The setting of thresholds is usually intended to be "conservative", such that in laboratory trials an allelic designation is rarely made when not correct. However, a reduction in one kind of error for a binary classification inevitably implies an increase in the other error, so that true alleles are wrongly designated as stutter. Moreover, this "conservative" policy does not necessarily favour defendants: broadly speaking, fewer alleles called from a crime scene epg means less information to identify donors, which in general helps defendants, but calling a peak as stutter rather than allelic can disfavour a defendant whose profile does not include that allele.

Most importantly, many laboratories have rules for deciding when the genotype of the major donor to a mixture can be confidently identified, referred to as deconvolution of a mixed profile. Because of masking, it is almost never possible to determine the genotype of any donor other than a clear major. Typically, deconvolution is not permissible if the two greatest DNA donors gave similar amounts of DNA to the sample, but as the discrepancy in the amounts of DNA increases, it becomes increasingly possible to make a confident inference of the genotype of the major donor. Choosing criteria for when the major genotype can be inferred with sufficient confidence is problematic, and the implications can be considerable, because the evidential weight attached to a single genotype deconvolved from a mixture can be many orders of magnitude greater than for the mixed profile.

Experts are capable of making fine judgements about the plausibilities of different possible states of nature (e.g. stutter/allele). We believe that in general they do this well, better than decision processes based on thresholds, and that their ability can be assessed by inter-expert comparisons in blind trials. We propose to exploit the ability of a trained human expert to accurately process complex information, by allowing them to specify a probability distribution for the genotype of a donor of interest to a complex DNA profile. Typically, this donor will be the source of the largest amount of DNA, excluding known donors. For example, if the genotype of the major donor is clear at most loci, but there is some ambiguity at one or two loci, this information can be fully captured by specifying a unique genotype (probability one) at the relevant loci, but encoding the ambiguous loci as probability distributions concentrated on the genotypes consistent with the epg. If the donor of interest is a minor donor, while the major donor genotype is known, then the assignment of probabilities to genotypes should allow for any uncertainty due to masking of alleles by the major profile. We can also specify a joint genotype distribution for more than one unknown contributor, from which marginal genotype distributions can be obtained.

The genotype of an individual at a short tandem repeat (STR) locus consists of an unordered pair from a set of possible alleles. We encode an uncertain genotype at a locus as a GPM, a symmetric matrix, with non-negative entries that sum to one. The number of rows and columns is the number of alleles recognised at the locus. The diagonal entries are the probabilities of the homozygote genotypes, and the $(i,j)$th and $(j,i)$th entries are each half the probability of an $ij$ heterozygote. GPMs can be used to encode epg information or information about an individual's genotype obtained from genotypes of their relatives.

## 3 Calculating single-locus LRs using GPMs

Consider two GPMs, $G^1$ and $G^2$, and suppose that we seek to compare the propositions:

$\eta_1$ : $G^1$ and $G^2$ represent the genotype of the same individual.

$\eta_0$ : $G^1$ and $G^2$ represent the genotypes of two unrelated individuals (who can have the same genotype).

In a standard setting, one GPM is from a CSP and may reflect uncertainty as discussed above, while the other encodes a reference profile that is usually assumed to be measured without error. However, our framework treats $G^1$ and $G^2$ in the same way, and hence allows for uncertainty in both profiles. An uncertain reference profile can arise if the individual is not available to give a good-quality DNA sample, and it has been obtained from a personal item such as a comb or toothbrush, or when it has been inferred from the genotypes of one or more relatives. In other situations there may be no reference profile, and the task is to assess whether the same individual has contributed to different CSPs.

We will initially assume that, under $\eta_0$, the two individuals are unrelated and have no coancestry (that is $F_{ST} = 0$, see below). The case that the two individuals have a specified relationship, or are unrelated but share coancestry for example due to population structure, is discussed below. Genotypes are regarded a priori as random draws from a GPM representing the population genotype distribution, which we denote B (for background). It is standard practice in forensic DNA analysis to assume independence of each individual's two alleles at a locus, which is called Hardy-Weinberg Equilibrium (HWE) [7]. In that case B can be expressed in the form $B = \mathbf{b}^T\mathbf{b}$ where $\mathbf{b}$ denotes a row vector of allele probabilities. Here, $^T$ denotes transpose, so $\mathbf{b}^T$ is a k x 1 column vector and B is k x k. In practice, the elements of $\mathbf{b}$ are relative frequencies obtained from a population database. There may be evidence suggesting a specific ethnic background of a donor of interest and hence an appropriate choice of B, or the weight of evidence may be assessed for multiple B matrices reflecting different possible ethnic backgrounds.

The LR conveying the weight of DNA evidence in support of $\eta_1$ relative to $\eta_0$ is the ratio of the probability of the evidence if $\eta_1$ is true, to its probability if $\eta_0$ holds [7].

**Theorem 1**

$$LR(\eta_1, \eta_0) = \sum_{ij} \frac{G^1_{ij} \cdot G^2_{ij}}{B_{ij}} \qquad (1)$$

where the subscript *ij* denotes the *ij*th entry of a matrix. The proof is in Appendix A1.

Each element of the numerator of (1) is the product of the probabilities that an *ij* genotype underlies each of samples 1 and 2. The denominator, $B_{ij}$ is the probability for an unknown, unprofiled individual to have genotype *ij*. In (1), the ratio of these terms is summed over all genotypes at the locus.

Here, we assume that only one population is relevant, so that the $B_{ij}$ apply to both contributors under $\eta_0$. Equation (1) holds approximately if the two individuals come from different populations each with their own background genotype frequency matrix: in this case the B that represents the alternative contributor under $\eta_0$ should be used. In Section (6) we generalise (1) to allow for the two individuals under $\eta_0$ to come from the same subpopulation of the population from which the $B_{ij}$ are obtained (coancestry correction).

In the special case that both $G^1$ and $G^2$ assign probability 1 to one genotype (spread over two entries of 0.5 for a heterozygote), (1) simplifies to $1/(2B_{ij})$ (using $B_{ij} = B_{ji}$), if genotype ij is specified by both $G^1$ and $G^2$ ($1/B_{ii}$ for a homozygote), and zero otherwise. This corresponds to the familiar result that the LR is the reciprocal of the population genotype probability, which is the special case of the match probability when $F_{ST} = 0$.

As an example, consider a locus with three known alleles $(a,b,c)$. Suppose we have a CSP from which an expert judges $P(aa)=0.5, P(ab)=0.5$, a reference profile $P(ab)=1$ (ie certainly ab), and $P(a)=0.5, P(b)=0.3, P(c)=0.2$. Then we have GPMs:

$$G1=\begin{pmatrix} 0.5 & 0.25 & 0 \\ 0.25 & 0 & 0 \\ 0 & 0 & 0 \end{pmatrix}, \; G2=\begin{pmatrix} 0 & 0.5 & 0 \\ 0.5 & 0 & 0 \\ 0 & 0 & 0 \end{pmatrix}, \; B=\begin{pmatrix} 0.25 & 0.15 & 0.10 \\ 0.15 & 0.09 & 0.06 \\ 0.10 & 0.06 & 0.04 \end{pmatrix}$$

and we may calculate the LR from (1) as

$$LR=\sum \frac{\begin{pmatrix} 0.5 & 0.25 & 0 \\ 0.25 & 0 & 0 \\ 0 & 0 & 0 \end{pmatrix} \cdot \begin{pmatrix} 0 & 0.5 & 0 \\ 0.5 & 0 & 0 \\ 0 & 0 & 0 \end{pmatrix}}{\begin{pmatrix} 0.25 & 0.15 & 0.10 \\ 0.15 & 0.09 & 0.06 \\ 0.10 & 0.06 & 0.04 \end{pmatrix}} = \sum \begin{pmatrix} 0 & 0.833 & 0 \\ 0.833 & 0 & 0 \\ 0 & 0 & 0 \end{pmatrix} = 1.666$$

(Note that matrix multiplication ( $\cdot$ ) and division are both element-wise. If the CSP had been found with certainty to be identical to the reference profile we would have

$$LR=\sum \frac{\begin{pmatrix} 0 & 0.5 & 0 \\ 0.5 & 0 & 0 \\ 0 & 0 & 0 \end{pmatrix} \cdot \begin{pmatrix} 0 & 0.5 & 0 \\ 0.5 & 0 & 0 \\ 0 & 0 & 0 \end{pmatrix}}{\begin{pmatrix} 0.25 & 0.15 & 0.10 \\ 0.15 & 0.09 & 0.06 \\ 0.10 & 0.06 & 0.04 \end{pmatrix}} = \sum \begin{pmatrix} 0 & 1.667 & 0 \\ 1.667 & 0 & 0 \\ 0 & 0 & 0 \end{pmatrix} = 3.333$$

which equals the standard result $LR=1/P(ab)=1/(2*0.5*0.3)=3.333$. The reader may wish to check that if the same probabilistically encoded matrix G1 is matched against a reference profile that is certainly aa then LR = 2.0.

## 4  LRs under specified relationships

A GPM for any individual can be used to generate GPMs for their parents and children, and subsequently other relatives. To see this, note that a GPM specifies an allele probability vector via its row or column sums (because GPMs are symmetric, these are the same). If **p₁** and **p₂** denote allele probability vectors for two parents, under HWE the GPM for their child can be written

$$G = (\mathbf{p_1}^T\mathbf{p_2} + \mathbf{p_2}^T\mathbf{p_1})/2. \qquad (2)$$

We ignore the possibility of mutation here, and we continue to assume $F_{ST} = 0$ (no coancestry). If there is no genotype information available for one parent, then **b**, the population allele probability vector, can be used instead:

**Figure 1**: *GPMs for relatives of a profiled individual.*

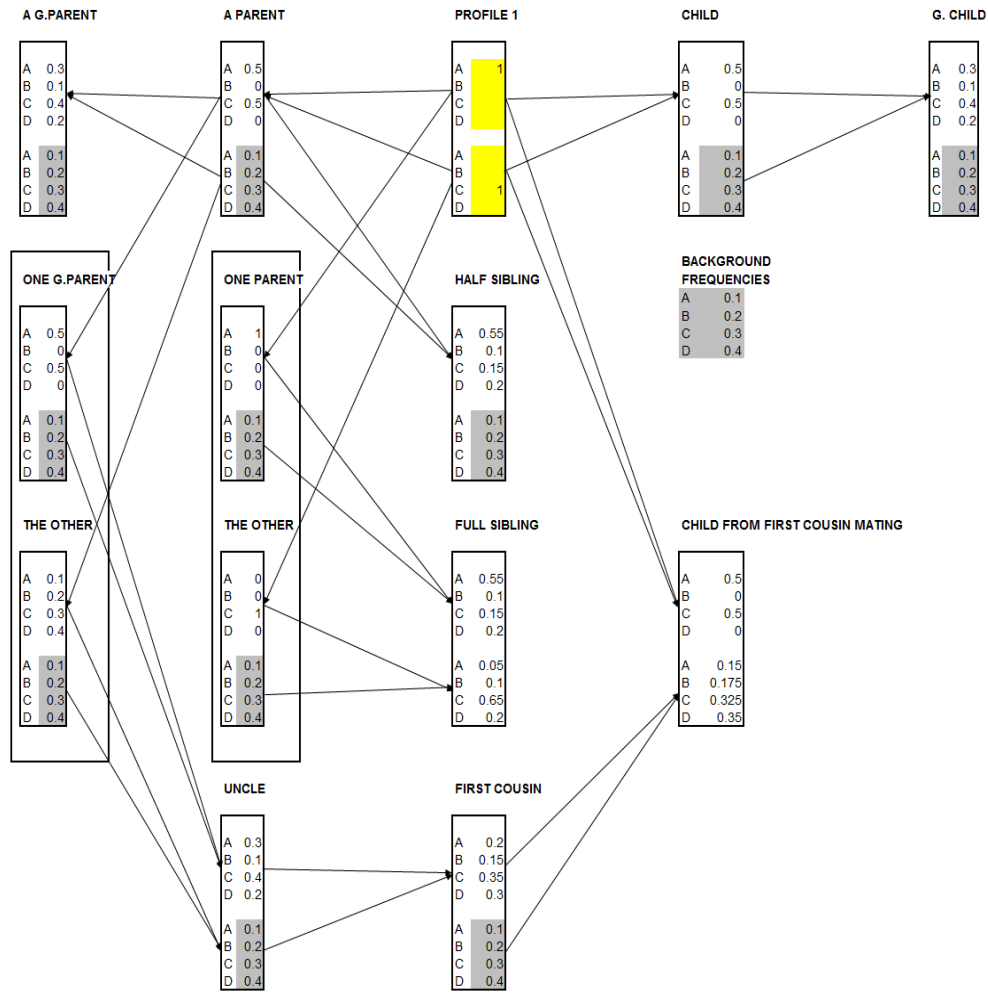

Under HWE, each GPM can be specified by two parental allele probability vectors, $p_1$ and $p_2$, which are given here for each individual at a four-allele locus. The GPM of a reference individual is specified (here Profile 1 has genotype AC, indicated by the $p_1$ and $p_2$ assignments highlighted in yellow). The population allele probability vector **b** is shaded in grey at each instance. For example, CHILD has one allele from Profile 1, which is equally likely to be A or C, and one allele chosen according to **b**. The diagram shows both the marginal GPM for a parental genotype ("A PARENT"), and also the joint GPMs of both parents ("ONE PARENT" and "THE OTHER", both included in a box). A similar situation applies to the grandparents through a specified parent.

$$G = (p^T b + b^T p)/2. \quad (3)$$

The GPM of a parent given a child's allele probability vector **c** is

$$G = (c^T b + b^T c)/2 \quad (4)$$

which is equivalent to (3), reflecting the symmetry of the parent-child relationship (and any unilineal relationship) when no other genotyped relatives are available. Given a GPM for an individual, the GPM for any unilineal relative can be obtained by applying the above two steps for every parent-child link separating the relatives (1. generate an allele frequency vector from the GPM; 2. use (3) or (4) to generate a new GPM). Explicit formulae for some common relationships are developed in Appendix A2.

Specifying the joint probability distributions for two parents is more complex, because their genotypes are dependent given their child's genotype. For simplicity, we focus only on a pair of relatives, one of whom is genotyped. Relationships among multiple individuals pose no problem in principle: Bayesian networks provide a good framework for propagating GPMs through arbitrary networks of related individuals, and GPMs can be assigned to any specified members of the network [8].

Figure 1 illustrates the derivation of GPMs via relatedness in some important special cases. Assuming HWE, these GPMs can always be expressed in terms of two allele probability vectors. This simpler representation is not always available for GPMs derived directly from an epg. In Figure 1 the genotype for Profile 1 is known with certainty, but we can also combine uncertainty from a noisy epg with uncertainty due to relatedness.

If G denotes a GPM for an individual, we will write R(G) to denote the GPM derived from G for a specified relative, using one or more instances of (3) or (4). Note that R(B) = B, since the probability distribution for an individual of unknown genotype is the same as for any of their relatives. In place of hypothesis $\eta_1$, we now contrast $\eta_0$ with

$\eta_R$ : the individual underlying $G^1$ is an R-relative of the individual underlying $G^2$.

**Theorem 2**

$$LR(\eta_R, \eta_0) = \sum_{ij} \frac{G^1_{ij} \cdot R(G^2_{ij})}{B_{ij}} \qquad (5)$$

See Appendix A3 for proof. Equation (5) simplifies to standard LR formulae for relatives (see for example Fung [9]) in the special case that the GPMs $G^1$ and $G^2$ assign probability 1 to a single genotype at each marker.

As an example consider the matrices G1, G2 and B from the previous section, but now ask whether the two samples come from full siblings. The derivation of a GPM for a specified relative of a profiled individual is described in appendix A2. From the final row of table A2-2 we find that

$$Sib(G2) = \left[ \begin{pmatrix} 0 & 0.5 & 0 \\ 0.5 & 0 & 0 \\ 0 & 0 & 0 \end{pmatrix} + \begin{pmatrix} 0.25 & 0.15 & 0.10 \\ 0.25 & 0.15 & 0.10 \\ 0 & 0 & 0 \end{pmatrix} + \begin{pmatrix} 0.25 & 0.25 & 0 \\ 0.15 & 0.15 & 0 \\ 0.10 & 0.10 & 0 \end{pmatrix} + \begin{pmatrix} 0.25 & 0.15 & 0.10 \\ 0.15 & 0.09 & 0.06 \\ 0.10 & 0.06 & 0.04 \end{pmatrix} \right] / 4$$

$$= \begin{pmatrix} 0.1875 & 0.2625 & 0.05 \\ 0.2625 & 0.0975 & 0.04 \\ 0.0500 & 0.0400 & 0.01 \end{pmatrix}$$

and we calculate the LR from (5) as

$$LR = \sum \frac{\begin{pmatrix} 0.5 & 0.25 & 0 \\ 0.25 & 0 & 0 \\ 0 & 0 & 0 \end{pmatrix} \cdot \begin{pmatrix} 0.1875 & 0.2625 & 0.05 \\ 0.2625 & 0.0975 & 0.04 \\ 0.0500 & 0.0400 & 0.01 \end{pmatrix}}{\begin{pmatrix} 0.25 & 0.15 & 0.10 \\ 0.15 & 0.09 & 0.06 \\ 0.10 & 0.06 & 0.04 \end{pmatrix}} = \sum \begin{pmatrix} 0.3750 & 0.4375 & 0 \\ 0.4375 & 0 & 0 \\ 0 & 0 & 0 \end{pmatrix} = 1.25$$

# 5  Mutation models.

In Figure 1, and in deriving (5), we assumed no mutation at the locus under consideration in the genetic lineage linking the two relatives. That assumption was for convenience only and is not required.

Mutations occur at forensic STR loci roughly once every 500 generations in males, and once every 2000 generations in females [10]. About 97% of mutations change the repeat count by one. Thus the stepwise mutation model, in which all mutations are between alleles that differ by one repeat unit, is incorrect but can provide a good approximation. The further simplifying assumption of symmetry (steps up as likely as steps down) may be adequate over small numbers of generations, but steps up are more common [10]. We introduce a mutation matrix M, whose *ij*th entry is the probability that a child receives a *j* allele given that the parental allele was an *i*. Thus 1-$M_{ii}$ is the total mutation rate for an *i* allele.

To compute the GPM of an individual, given the GPM of their relative and allowing for mutation in parent-child transmissions, we proceed as described in Section 4 and illustrated in Figure 1 except that a parental allele probability vector **p** should be replaced with **p**M when computing the child's GPM. The population allele probability vector **b** is assumed to be the same across generations. This implies **b**M = **b**, which holds at least approximately for realistic M. Theorem 2 continues to hold if R is replaced with $R_M$, denoting the transformation of a GPM due to relatedness allowing for mutation.

# 6  Population structure and coancestry

Even if two individuals have no close relatedness, they may share common ancestors a handful of generations in the past, so that their genotypes are positively correlated and the independence assumption underpinning Theorems 1 and 2 is not valid. The simplest scenario in which coancestry is important is the case that, under $\eta_0$, the individuals whose genotypes underlie $G^1$ and $G^2$ both come from the same subpopulation of the population from which **b** has been obtained. A dependence between the genotypes arises because the genotype of either one of them is informative about the local allele probabilities which alters the probability distribution for the other individual. A mathematical model has been developed for such dependence, sometimes called the Balding-Nichols conditional model [7]. It is the basis of equation 4.10 in the NRCII report [11]. It assumes in effect that the local allele probability vector has a Dirichlet distribution with mean **b** and variance determined by the population genetics parameter $F_{ST}$ (sometimes called $\theta$). Theorems 1 and 2 assume $F_{ST}$ = 0, corresponding to zero variance.

The details of the Balding-Nichols model are not important here. All we need is notation for conditional genotype probabilities. B still specifies the marginal GPM of an unknown individual, but the GPM for a second unknown individual requires a GPM conditional on the genotype, say *ij*, of the first individual; we will denote this conditional GPM $B_{|ij}$. Thus $B_{pq|ij}$ is the $pq$ entry of $B_{|ij}$, the probability that an unknown individual has genotype $pq$ given the observation of an unrelated individual in the same subpopulation with genotype $ij$. See Appendix A3 for derivation of the LR which we now state, corresponding to Theorem 2 which allows for relatedness and mutation but now further extended to allow for coancestry between the two individuals under $\eta_0$:

$$LR(\eta_R, \eta_0) = \frac{\sum_{ij} \frac{G_{ij}^1 \cdot R_M(G_{ij}^2)}{B_{ij}}}{\sum_{ij} G_{ij}^1 \cdot H_{ij}^2} \quad ; \quad H_{ij}^2 = \sum_{pq} \frac{G_{pq}^2 \cdot B_{pq|ij}}{B_{pq}} \qquad (6)$$

The numerator of (6) equals (5), while the denominator is the "GPM subpopulation correction factor". It reduces to 1 when $F_{ST} = 0$, in which case $B_{|ij} = B$.

Equation (6) is our most general expression for the LR, and is valid for any choice of conditional probabilities $B_{pq|ij}$, including the Balding-Nichols subpopulation correction [7: S5.3.1.2]. Note however that the familial formulae given in Appendix 2 assume HWE which does not apply under the Balding-Nichols model. For these calculations GPMDNA implements the beta-binomial sampling formula underlying the Balding-Nichols model [7: S7.2] and also for the $R_M$ matrix (see Appendix A4).

# 7 Encoding an epg as a GPM

## 7.1 Using a statistical model

A number of statistical models and software for the evaluation of low-template DNA (LTDNA) profiles have been published in recent years [2, 12, 13,14]. LRs are calculated by summing over the possible genotypes of unknown contributors, using explicit models of processes such as dropout, dropin, stutter and degradation [15,16]. Some degree of expert judgement is still required in processing the epg into input data for the software, for example in encoding apparent artefacts.

In contrast, our approach uses the epg to infer contributor genotype(s) as a first step, but since the recent statistical models each imply a probability distribution for these genotypes there is no conflict. Indeed the statistical models can be used to derive contributor GPMs from an epg. However, here we emphasise that expert judgement can also form a valid basis for deriving GPMs: we return to discussing the advantages of our approach in the Conclusion.

In the special case that $G^1$ specifies a unique genotype (a reference profile), encoding a full GPM $G^2$ is not required since only the entries corresponding to the genotype specified by $G^1$ are used in (6). However, epgs should be encoded 'blind', without knowledge of the subsequent comparisons, and so a full GPM is required. Here, we consider the problem of encoding epgs in a range of common situations. The subsequent calculation of LRs using (1), (5) or (6) is a straightforward computation. We noted above that statistical algorithms are available for inferring genotypes from an epg, we here assume subjective deconvolution by a trained expert.

## 7.2 Expert assessment

When assigning genotype probabilities, the expert makes assessments of the number of significant contributors and the relative amounts of DNA from each of them, based on the observed peaks, particularly their heights. Note that the expert can choose to specify a GPM only for the major contributor, or for the two most important contributors, without drawing inferences about the genotypes of lesser contributors. There is no requirement to assess the total number of contributors of DNA to the sample: this number is of little importance when some individuals contribute negligible amounts of DNA. The GPM assessments should take into account the genotypes of known contributors, and allow for stochastic phenomena such as variability in peak heights, as well as possible drop out and drop in. When there is uncertainty about the genotype of a contributor, for example if there is a single epg peak that may represent a homozygote genotype or a heterozygote with one allele dropped out, then the expert may also take into account a vector **b** of background allele probabilities for unobserved alleles.

The following examples illustrate, for a range of common situations, how an expert may take these factors into account. An encoding scheme and shorthand notation can facilitate this task, see Appendix A4 for further discussion.

### 7.2.1 One contributor of interest

If the contributor of interest is the major contributor of DNA to the sample, and allelic peaks are well above noise levels, encoding is reasonably straightforward. At loci where there are one or two peaks much higher than the others, the expert can assign high probability to a single genotype, whereas at other loci there may be more ambiguity about the major peaks, requiring probability judgements based on peak heights and possibly also **b**.

*Example 1: Single contributor, allowing for dropin and dropout*

Consider two peaks with approximate height ratio of A:B = 2:1 at a locus with only 3 alleles: A, B, and C. In Table 1 we show three schemes for the expert to encode the epg information, all assuming that there is at least one A allele present, while the peak at B may be allelic or a dropin. Assuming the allele probabilities are known, the three schemes are in order of increasing complexity, with 0, 1 and 2 parameters. Row (i) corresponds to coding the genotype as AF (see discussion above): it is assumed that because the peak at B could be a dropin, the epg conveys no information about the second allele, which may have dropped out. Thus the allele probabilities are given by **b**. This corresponds to some current practice but it ignores available information: the relative peak heights.

| | | Genotypes | | |
|---|---|---|---|---|
| | | AA | AB | AC |
| options | i | $b_A$ | $b_B$ | $b_C$ |
| | ii | $\Gamma$ | $1-\Gamma$ | 0 |
| | iii | $\Gamma(1-\Delta)+\Gamma\Delta b_A$ | $(1-\Gamma)+\Gamma\Delta b_B$ | $\Gamma\Delta b_C$ |

Table 1: Genotype probability allocations for options (i) – (iii) discussed in the text.

In row (ii), the expert decides to ignore the possibility that a dropout has occurred but considers that the B allele may be a dropin, with probability $\Gamma$. Then the genotype is AA with probability $\Gamma$ and AB with probability $1-\Gamma$. Row (iii) represents an interpolation between row (i) (case $\Delta=1$) and row (ii) (case $\Delta=0$). The value of $\Delta$ reflects the probability of dropout: $\Delta=0$ implies zero dropout and so the AC genotype is impossible, whereas $\Delta=1$ implies a high probability of dropout. The probability of dropout may be inferred subjectively by the encoder based on the observed peak heights, or else could be inferred using an established algorithm, for example [19]. A GPM resulting from a particular instance of the row (iii) coding is shown in Table 2.

|   | A | B | C |
|---|---|---|---|
| A | 0.220 | 0.32 | 0.07 |
| B | 0.320 | 0 | 0 |
| C | 0.07 | 0 | 0 |

Table 2: EPG encoding from row (iii) of Table 1 with $b_A=0.1, b_B=0.2, b_C=0.7, \Gamma=0.4, \Delta=0.5$

We illustrate LRs for the GPM of Table 2 given reference profiles respectively AA, AB, and AC

$$AA: LR = \sum \begin{pmatrix} 1 & 0 & 0 \\ 0 & 0 & 0 \\ 0 & 0 & 0 \end{pmatrix} \cdot \begin{pmatrix} 0.22 & 0.32 & 0.07 \\ 0.32 & 0 & 0 \\ 0.07 & 0 & 0 \end{pmatrix} / \begin{pmatrix} 0.01 & 0.02 & 0.07 \\ 0.02 & 0.04 & 0.14 \\ 0.07 & 0.14 & 0.49 \end{pmatrix} = \sum \begin{pmatrix} 22 & 0 & 0 \\ 0 & 0 & 0 \\ 0 & 0 & 0 \end{pmatrix} = 22$$

$$AB: LR = \sum \begin{pmatrix} 0 & 0.5 & 0 \\ 0.5 & 0 & 0 \\ 0 & 0 & 0 \end{pmatrix} \cdot \begin{pmatrix} 0.22 & 0.32 & 0.07 \\ 0.32 & 0 & 0 \\ 0.07 & 0 & 0 \end{pmatrix} / \begin{pmatrix} 0.01 & 0.02 & 0.07 \\ 0.02 & 0.04 & 0.14 \\ 0.07 & 0.14 & 0.49 \end{pmatrix} = \sum \begin{pmatrix} 0 & 8 & 0 \\ 8 & 0 & 0 \\ 0 & 0 & 0 \end{pmatrix} = 16$$

$$AC: LR = \sum \begin{pmatrix} 0 & 0 & 0.5 \\ 0 & 0 & 0 \\ 0.5 & 0 & 0 \end{pmatrix} \cdot \begin{pmatrix} 0.22 & 0.32 & 0.07 \\ 0.32 & 0 & 0 \\ 0.07 & 0 & 0 \end{pmatrix} / \begin{pmatrix} 0.01 & 0.02 & 0.07 \\ 0.02 & 0.04 & 0.14 \\ 0.07 & 0.14 & 0.49 \end{pmatrix} = \sum \begin{pmatrix} 0 & 0 & 0.5 \\ 0 & 0 & 0 \\ 0.5 & 0 & 0 \end{pmatrix} = 1$$

### 7.2.2 Two contributors of interest

It is difficult in general to specify a joint genotype distribution for multiple contributors, because for example the genotype probabilities for a second contributor will depend on genotype assignments for the first contributor. However, it is feasible in some settings.

If the differences in peak heights between the two contributors are large, it may be possible to infer the genotype of the major with certainty, in which case the task reduces to encoding the genotype of a minor with known major (see Example 2 below).

Our approach can also be used when the two main contributors can to some extent be distinguished through peak heights, but there remains some uncertainty about the genotype of the major at some loci, and considerable uncertainty about the minor at most loci. We have outlined above how to specify a GPM for the major contributor. A GPM for a minor can be similarly derived conditional on each possible genotype for the major. A marginal GPM for the minor is then obtained by summing over the major contributor genotypes, weighted by their probabilities. Then the GPMs for major and minor contributors can both be used to compute LRs as outlined above, for example to search against databases.

A joint genotype distribution can be represented, if required, as a 4-dimensional matrix of allele probabilities, or more conveniently as a 2-dimensional matrix of genotype probabilities as in Table 3.

*Example 2 : Two distinguishable contributors.*

Consider three peaks with heights A:B:C = 5:5:2. Suppose that the expert decides that the major is AB, and the minor includes at least one C allele, while the other is equally likely to be masked by any of A, B or C, so that the joint genotype probabilities are as shown in Table 3. The corresponding GPMs for major and minor contributors are shown in Table 4 and Table 5.

|  |  | Major contributor |  |  |  |  |  |  | Minor Marginal |
|---|---|---|---|---|---|---|---|---|---|
|  |  | AA | AB | AC | BB | BC | CC |  |  |
| Minor contributor | AA |  |  |  |  |  |  |  |  |
|  | AB |  |  |  |  |  |  |  |  |
|  | AC |  | 0.33 |  |  |  |  |  | 0.33 |
|  | BB |  |  |  |  |  |  |  |  |
|  | BC |  | 0.33 |  |  |  |  |  | 0.33 |
|  | CC |  | 0.33 |  |  |  |  |  | 0.33 |
| Major Marginal |  |  | 1 |  |  |  |  |  |  |

Table 3: Encoding of Example 2

|  | A | B | C |
|---|---|---|---|
| A | 0 | 0.5 | 0 |
| B | 0.5 | 0 | 0 |
| C | 0 | 0 | 0 |

Table 4: GPM corresponding to Major in Example 2.

|  | A | B | C |
|---|---|---|---|
| A | 0 | 0 | 0.167 |
| B | 0 | 0 | 0.167 |
| C | 0.167 | 0.167 | 0.333 |

Table 5: GPM corresponding to Minor in Example 2.

Example 3 : Two distinguishable contributors, neither certain

Consider four peaks with heights A:B:C:D = 5:3:3:2. The expert may decide that the four alleles are A, B, C and D; that the A and D peaks are major and minor alleles respectively, while B and C are each equally likely to be major and minor alleles[1]. These assignments are shown in Table 6, while the corresponding marginal GPMs are shown in Table 7 and Table 8. Note that the joint genotype probabilities do not correspond to the product of the marginal probabilities. For example, there is uncertainty about the origin of the peaks at both B and C, but resolving one of them determines the other.

---

1  In Example 3 since four alleles have been identified dropout may be ignored, and since the same four alleles are identified in either scenario, differences in their relative background frequencies do not affect the assignment.

|  |  | Major contributor |  |  |  | Minor Marginal |
|---|---|---|---|---|---|---|
|  |  | AA | AB | AC | AD |  |
| Minor contributor | AD |  |  |  |  |  |
|  | BD |  |  | 0.5 |  | 0.5 |
|  | CD |  | 0.5 |  |  | 0.5 |
|  | DD |  |  |  |  |  |
| Major Marginal |  |  | 0.5 | 0.5 |  |  |

Table 6: Encoding of Example 3

|  | A | B | C | D |
|---|---|---|---|---|
| A | 0 | 0.25 | 0.25 | 0 |
| B | 0.25 | 0 | 0 | 0 |
| C | 0.25 | 0 | 0 | 0 |
| D | 0 | 0 | 0 | 0 |

Table 7: GPM corresponding to Major in Example 3.

|  | A | B | C | D |
|---|---|---|---|---|
| A | 0 | 0 | 0 | 0 |
| B | 0 | 0 | 0.25 | 0.25 |
| C | 0 | 0.25 | 0 | 0 |
| D | 0 | 0.25 | 0 | 0 |

Table 8: GPM corresponding to Minor in Example 3.

*Example 4: Two indistinguishable contributors*

Consider three peaks with heights A:B:C: = 5:4:3. Suppose that the expert decides there is insufficient information to distinguish between contributors, and that alleles A, B and C are all present, while the fourth allele is likely to be another A but may be another B.

Using her expert judgement (taking into account the peak heights and background frequencies), suppose she assigns probabilities as to which alleles are present as follows:
 P(AABC) = 0.9
 P(ABBC) = 0.1
We note that given any four alleles the six possible genotype pairs into which they can be arranged are equally likely. Writing *p* and *q* for two constants in the ratio 9:1, Table 9 identifies all combinations of two contributors that can result in either AABC or ABBC being observed, and assigns probability *p* to each genotype pair containing alleles AABC, and *q* to each pair containing ABBC (taking into account factors of two for the two possible orderings of a heterozygote). The row sums (= column sums) specify the marginal genotype distribution, and equating the overall sum to one specifies *p* and *q* (*12p + 12q = 1; p/q = 9*; hence *p* = 9/120, *q* = 1/120).

|  |  | First contributor |  |  |  |  | Marginal |
|---|---|---|---|---|---|---|---|
|  |  | AA | AB | AC | BB | BC |  |
| Second contributor | AA |  |  |  |  | 2p | 2p |
|  | AB |  |  | 4p |  | 4q | 4p+4q |
|  | AC |  | 4p |  | 2q |  | 4p+2q |
|  | BB |  |  | 2q |  |  | 2q |
|  | BC | 2p | 4q |  |  |  | 2p+4q |
|  | Marginal | 2p | 4p+4q | 4p+2q | 2q | 2p+4q |  |

Table 9: Joint genotype probabilities for two indistinguishable contributors in Example 4.

|  | A | B | C |
|---|---|---|---|
| A | 18q = 0.15 | 20q = 0.17 | 19q = 0.16 |
| B | 20q = 0.17 | 2q = 0.02 | 11q = 0.09 |
| C | 19q = 0.16 | 11q = 0.09 | 0 |

Table 10: Marginal GPM corresponding to the final row (or column) of Table 9, using p = 9q and (q = 1/120).

The GPM in Table 10 (restating the final row and column of Table 9) represents the probability distribution for the genotype of a contributor to the mixture: since the expert does not attempt to deconvolve the mixture, the GPMs are the same for each contributor. If searched against a database of GPMs, a non-zero LR would be returned for any GPM that assigns a non-zero probability to any of the genotypes AA, AB, AC, BB or BC. The LR computed using GPMs for two non-deconvolved mixtures measures the evidence for the proposition that the mixtures have a contributor in common, without specifying the genotype of the common contributor.

For example, the LR for the proposition that the major contributor in example 3 is one of the contributors to example 4 is obtained by using the GPMs of Table 7 and Table 10 in Theorem 1, which gives $0.085/b_A b_B + 0.080/b_A b_C$ which equals 16.5 in the special case that $b_A = b_B = b_C = 0.1$.

# 8 Conclusion

We propose computing LRs to evaluate poor-quality DNA evidence (due to low-template, degradation, and/or multiple donors) using a probability distribution for the genotype(s) of donor(s). While the donor genotypes are strictly nuisance variables and it is possible to compute LRs without inferring them, we see several advantages to explicitly using genotype probability distributions for unknown or contested contributors. Firstly, this is an intuitive quantity for which an expert can reasonably assign probabilities, thus taking advantage of their expert judgment. Secondly, summarising the information in a complex epg in terms of one or more genotype distributions allows rapid searching of even complex profiles against large databases, as well as the comparison of two complex profiles, perhaps from different crime scenes, to assess hypotheses about common contributors. A final advantage is that for an individual with a known genotype, probability distributions for the genotypes of their relatives are readily computed, allowing us to assess hypotheses about relatedness of contributors to different profiles even when one or both profiles are complex. We believe that our approach can assist both in rapid identification of intelligence leads and in evaluating evidence within a judicial process.

Although the use of "subjective" encoding based upon expert opinion may seem problematic, there is now a substantial literature on the elicitation of expert opinion in a subjective Bayesian methodology [17, 18]. Moreover the performance of experts can be tested in blind trials and through assessments of between-expert agreement. The expert can make conservative assessments, for example by assigning more weight to background frequencies for noisy profiles. Initially the benefits of the method will be in its use as an intelligence tool; providing additional leads to an investigation. New and historical epgs that could not be used conventionally can now be searched and stored to the advantage of ongoing and "cold case" investigations. We envisage crime datasets made up of probabilistic epg encodings that would be a subset of, or complementary to, conventional national crime databases. Any possible matches suggested could then be subject to conventional examination. The use of the method within judicial systems would, we hope, follow. There are clear advantages in being able to give evidential strengths rather than a narrative along the lines of "could not be excluded as a donor to...".

# A1 Theorem 1

Given epgs $E^1$ and $E^2$ and hypotheses

$\eta_1$ : $G^1$ and $G^2$ represent the genotype of the same individual

$\eta_0$ : $G^1$ and $G^2$ represent the genotypes of two unrelated individuals

the likelihood ratio (LR) is

$$LR(\eta_1, \eta_0) = \frac{P(E^1, E^2 \mid \eta_1)}{P(E^1, E^2 \mid \eta_0)} \tag{A1-0}$$

We write $S^1$ and $S^2$ for the genotypes of the contributors to $E^1$ and $E^2$. Under $\eta_1$ we have $S^1 = S^2$ and the numerator is obtained by summing over the two alleles of the genotype. Under $\eta_0$, we marginalize over the four alleles $A_i A_j A_p A_q$ where $S^1 = A_i A_j : i \leq j$ and $S^2 = A_p A_q : p \leq q$. Then

$$LR = \frac{\sum_{ij} P(E^1, E^2 \mid S^1 = S^2 = A_i A_j, \eta_1) \cdot P(S^1 = S^2 = A_i A_j \mid \eta_1)}{\sum_{pqij} P(E^1, E^2 \mid S^1 = A_i A_j, S^2 = A_p A_q, \eta_0) \cdot P(S^1 = A_i A_j, S^2 = A_p A_q \mid \eta_0)} \tag{A1-1}$$

We now make the following assumptions

1. In the numerator, $E^1$ and $E^2$ are assumed independent given the underlying genotype $A_i A_j$ so that $P(E^1, E^2 \mid S^1 = S^2 = A_i A_j, \eta_1)$ becomes $P(E^1 \mid S^1 = A_i A_j) \cdot P(E^2 \mid S^2 = A_i A_j)$.

2. $P(S^1 = S^2 = A_i A_j \mid \eta_1)$ is the background genotype probability $B_{ij}$.

3. In the denominator $P(E^1, E^2 \mid S^1, S^2, \eta_0) = P(E^1 \mid S^1 = A_i A_j) \cdot P(E^2 \mid S^2 = A_p A_q)$ again assuming $E^1$ and $E^2$ are independent given the genotypes.

4. $P(S^1 = A_i A_j, S^2 = A_p A_q \mid \eta_0) = P(S^1 = A_i A_j \mid \eta_0) \cdot P(S^2 = A_p A_q \mid S^1 = A_i A_j, \eta_0) = B_{ij} \cdot B_{pq \mid ij}$ using the notation for $\mathbf{B}_{\mid ij}$ introduced in Section 6.

Then

$$LR = \frac{\sum_{ij} P(E^1 \mid S^1 = A_i A_j) \cdot P(E^2 \mid S^2 = A_i A_j) \cdot B_{ij}}{\sum_{ij} \left[ P(E^1 \mid S^1 = A_i A_j) \cdot B_{ij} \cdot \sum_{pq} P(E^2 \mid S^2 = A_p A_q) \cdot B_{pq \mid ij} \right]} \tag{A1-2}$$

This equation contains terms of the form $P(E \mid S = A_i A_j)$. This is appropriate when considering explicit models of the generation of epgs from profiles. We now use Bayes' theorem to write the LR in terms of the form $G_{ij} = P(S = A_i A_j \mid E)$, which correspond to entries of GPM matrices. By Bayes' theorem:

$P(E \mid S = A_i A_j) = G_{ij} P(E)/B_{ij}$ where $P(E)$ is the *prior probability* of the epg $E$. Substituting this expression into (A2-2) we get

$$LR = \frac{\sum_{ij} \frac{G_{ij}^1 P(E^1)}{B_{ij}} \cdot \frac{G_{ij}^2 P(E^2)}{B_{ij}} \cdot B_{ij}}{\sum_{ij} \left[ \frac{G_{ij}^1 P(E^1)}{B_{ij}} \cdot B_{ij} \cdot \sum_{pq} \frac{G_{pq}^2 P(E^2)}{B_{pq}} \cdot B_{pq|ij} \right]}$$

And simplifying

$$LR = \frac{\sum_{ij} \frac{G_{ij}^1 \cdot G_{ij}^2}{B_{ij}}}{\sum_{ij} \left[ G_{ij}^1 \cdot \sum_{pq} \frac{G_{pq}^2 \cdot B_{pq|ij}}{B_{pq}} \right]} \quad (A1\text{-}3)$$

If $S^1$ and $S^2$ are independent then $B_{pq|ij} = B_{pq}$ ) and the denominator simplifies to unity (Since $\sum_{ij} G_{ij}^1 = \sum_{ij} G_{ij}^2 = 1$ ), which establishes Theorem 1.

In the general case $B_{pq|ij} \neq B_{pq}$ (A1-3) should be used.

# A2 Relatedness

In this section we derive formulae equivalents to Figure 1 section 4.

We define $Rel_R(A_p A_q)_{ij}$ to be the probability that an R-relative of an individual with genotype $A_p A_q$ has genotype $A_i A_j$.

The function $Rel_R(A_p A_q)_{ij}$ is calculated on the following assumptions:

(a) Mendelian segregation with independent assortment.

(b) Genes contributed by unknown individuals are drawn from the background allele probability vector **b**. (Ie HW Equilibrium)

## A2.1 Degree 1

We consider first the parent-child relationship under these assumptions. We will show that the GPMs for "Child of X" and "Parent of X" are the same.

### A2.1.1 **Child**

We seek an expression for $Rel_{Child}(A_p A_q)_{ij}$ (i.e. component (i, j) of the matrix representing $Rel_{Child}(A_p A_q)$) or in words: "The probability that the child of an individual with genotype $A_p A_q$ is $A_i A_j$"

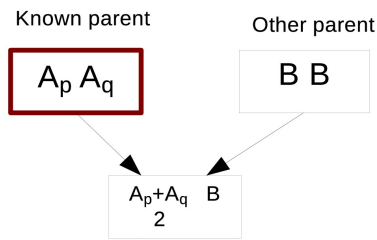

The child of an individual with genotype $A_p A_q$ will inherit $A_p$ or $A_q$, each with probability 0.5. The inherited allele may be written as the vector[2] $(\delta_{ip}+\delta_{iq})/2$

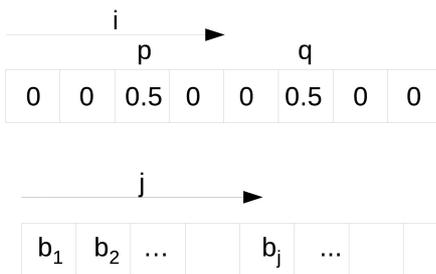

The other parent contributes an allele distributed according to the background probabilities **b**. The matrix $Rel_{Child}(A_p A_q)$ is then a symmetric product of these allele vectors:

---

[2] Where $\delta_{ij}$ is the Kronecker delta symbol, defined to be equal to 1 if $i = j$ and 0 otherwise.

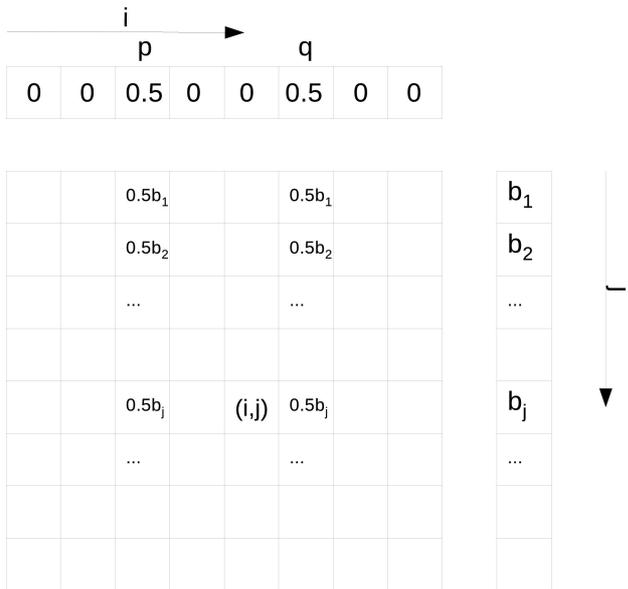

and it can be seen that the general term ((i,j) in the diagram) is

$$Rel_{Child}(A_p A_q)_{ij} = \frac{(\delta_{ip}+\delta_{iq})}{2} b_j \qquad (A2\text{-}1)$$

### A2.1.2 Parent

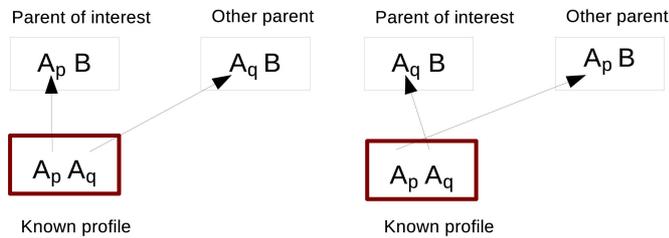

In the case of a parent there are two cases to consider: The parent contributed either $A_p$ or $A_q$, with the other allele drawn from **b**. Each of these cases may be constructed as above, and then averaged. (In the diagrams, arrows represent deductions, rather than gene flow).

$$Rel_{Parent}(A_p A_q)_{ij} = \frac{1}{2}(\delta_{ip} b_j + \delta_{iq} b_j) = \frac{(\delta_{ip}+\delta_{iq})}{2} b_j \qquad (A2\text{-}2)$$

It can be seen the expressions for a parent and a child are identical, and we term both these relationships *Degree 1 or* $D1()$.

## A2.2 Degree n

We term an *n*th generation descendant or ancestor (i.e. (n-2) greats grandchild or grandparent) a *Degree n* relative $Dn()$. Applying the construction for degree 1 repeatedly yields

$$Rel_{Dn}(A_p A_q)_{ij} = \frac{(\delta_{ip}+\delta_{iq}+(2^n-2)b_i)}{2^n} b_j \tag{A2-3}$$

**As special cases, we have the formulae for the D2 and D3 cases:**

$$Rel_{D2}(A_p A_q)_{ij} = \frac{(\delta_{ip}+\delta_{iq}+2b_i)}{4} b_j \tag{A2-4}$$

$$Rel_{D3}(A_p A_q)_{ij} = \frac{(\delta_{ip}+\delta_{iq}+6b_i)}{8} b_j \tag{A2-5}$$

### A2.3 Sibling

The full sibling relationship is the simplest example of a *collateral bilineal* relationship (the individuals have common ancestors via both parents), which may be constructed as follows:

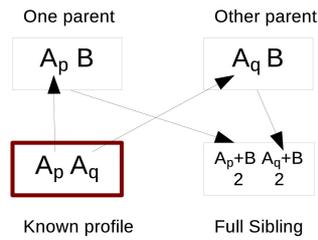

$$Rel_{Sib}(A_p A_q)_{ij} = \frac{1}{2}(b_i+\delta_{pi}) \cdot \frac{1}{2}(b_j+\delta_{qj}) \tag{A2-6}$$

NB in this case adding the alternative term obtained by interchanging *p* and *q* is equivalent to adding the symmetric term obtained by interchanging *i* and *j*, so it is omitted. This is not always true for bilineal relationships.

In the general case where the profile is represented by the GPM **G**

$$Rel_{Sib}(\boldsymbol{G})_{ij} = \sum_{pq} G_{pq} \cdot \frac{1}{2}(b_i+\delta_{pi}) \cdot \frac{1}{2}(b_j+\delta_{qj})$$

which may be simplified to

$$Rel_{Sib}(\boldsymbol{G})_{ij} = \frac{1}{4}(b_i b_j + b_i \sum_p G_{pj} + b_j \sum_q G_{iq} + G_{ij}) \tag{A2-7}$$

The LR for $A_p A_q$ and $A_r A_s$ being siblings (versus unrelated) is from Theorem 2)

which simplifies to

$$LR(A_p A_q, Rel_{Sib}(A_r A_s)) = \frac{1}{8}\left[\left(1+\frac{\delta_{pr}}{b_p}\right)\left(1+\frac{\delta_{qs}}{b_q}\right)+\left(1+\frac{\delta_{qr}}{b_q}\right)\left(1+\frac{\delta_{ps}}{b_p}\right)\right] \qquad (A2\text{-}8)$$

This formula reproduces all seven cases given in Table (A2-1) below, in agreement with published formulae, e.g. [13] at p137, Table 4.10.

| P1 | P2 | LR (Full Sibling/Unrelated) |
|---|---|---|
| xx | xx | $(1+b_x)^2 / 4 b_x^2$ |
| xx | xy | $(1+b_x)/4 b_x$ |
| xx | yy | $1/4$ |
| xx | yz | $1/4$ |
| xy | xy | $(1+b_x+b_y+2 b_x b_y)/(8 b_x b_y)$ |
| xy | xz | $(1+2 b_x)/8 b_x$ |
| xy | zw | $1/4$ |

*Table (A2-1): Sibling Likelihood Ratio*

## A2.4 Tables

| Relationship | Components of $Rel(A_p A_q)_{ij}$ | Vector notation: GPM of relative, given GPM $X$ with marginal $x$, background $B = b^T b$ |
|---|---|---|
| D1 | $(\delta_{ip}+\delta_{iq}) b_j / 2$ | $(x^T b + b^T x)/2$ |
| Dn | $(\delta_{ip}+\delta_{iq}+(2^n-2)b_i) b_j / 2^n$ | $(x^T b + b^T x + (2^n-2) B)/2^n$ |
| Full Sibling | $(b_i+\delta_{pi})\cdot(b_j+\delta_{qj})/4$ | $(X + x^T b + b^T x + B)/4$ |

*Table (A2-2): Relationship formulae*

| D1 | Parent, Child |
|---|---|
| D2 | Grandparent, Grandchild, Uncle, Nephew, Half-sibling |
| D3 | Great-grandparent, Great-grandchild, Great-aunt, Great-niece, First Cousin |

*Table (A2-3): Examples of relationships*

## A3 Theorem 2

Given an individual with GPM $\mathbf{G}$, the GPM $\mathbf{G^R}$ of a relative of that individual may be expressed as a function of $\mathbf{G}$ : $\mathbf{G^R} = R(\mathbf{G})$ where function $R()$ depends on the particular relationship, e.g. we may have $\mathbf{G^R} = Child(\mathbf{G})$, $\mathbf{G^R} = Sib(\mathbf{G})$ etc.

We now show how the likelihood ratio may be expressed in terms of $R()$.

We consider two GPMs $G^1$ and $G^2$ representing epgs $E^1$ and $E^2$ respectively. We consider the LR for the following hypotheses:

$\eta_R$ : the individual underlying $G^1$ is an R-relative of the individual underlying $G^2$.

$\eta_0$ : $G^1$ and $G^2$ come from unrelated individuals

Again we write $S^1$ and $S^2$ for the (unknown) genotypes of the contributors to $E^1$ and $E^2$. And $Rel_R(A_p A_q)_{ij}$ is defined as in Appendix A2. Then under $\eta_R$ the GPM for the relative $S^1$ is given by

$$R(\mathbf{G^2})_{ij} = \sum_{pq} G^2_{pq} Rel_R(A_p A_q)_{ij} \tag{A3-1}$$

Proceeding as in A2 we define the likelihood ratio for a familial match

$$LR = \frac{P(E^1, E^2 \mid \eta_R)}{P(E^1, E^2 \mid \eta_0)} \tag{A3-2}$$

First, marginalize both numerator and denominator over the exclusive and exhaustive set of alleles $A_p A_q A_i A_j$ where $S^1 = A_i A_j : i \leq j$ and $S^2 = A_p A_q : p \leq q$ :

$$LR = \frac{\sum_{pqij} P(E^1, E^2 \mid S^1 = A_i A_j, S^2 = A_p A_q, \eta_R) \cdot P(S^1 = A_i A_j, S^2 = A_p A_q \mid \eta_R)}{\sum_{pqij} P(E^1, E^2 \mid S^1 = A_i A_j, S^2 = A_p A_q, \eta_0) \cdot P(S^1 = A_i A_j, S^2 = A_p A_q \mid \eta_0)}$$

We proceed as in Appendix A1 with regard to the denominator.

In the numerator we have $P(E^1, E^2 \mid S^1, S^2, \eta_R) = P(E^1 \mid S^1, S^2, \eta_R) \cdot P(E^2 \mid E^1, S^1, S^2, \eta_R)$. Since under $\eta_R$ the sources of $S^1$ and $S^2$ are assumed to be related, it is *not* true that $E^1$ is independent of $S^2$, nor that $E^2$ is independent of $S^1$ and $E^1$. However it is clear that if $S^1$ were known, knowledge of $S^2$ or $E^2$ would not add any new information about $E^1$. Therefore $P(E^1 \mid S^1, S^2, \eta_R) = P(E^1 \mid S^1, \eta_R)$ and similarly $P(E^2 \mid E^1, S^1, S^2, \eta_R) = P(E^2 \mid S^2, \eta_R)$. Then

$$LR = \frac{\sum_{pqij} P(E^1 \mid S^1 = A_i A_j) \cdot P(E^2 \mid S^2 = A_p A_q) \cdot P(S^1 = A_i A_j \mid S^2 = A_p A_q, \eta_R) \cdot P(S^2 = A_p A_q \mid \eta_R)}{\sum_{pqij} P(E^1 \mid S^1 = A_i A_j) \cdot P(E^2 \mid S^2 = A_p A_q) \cdot P(S^2 = A_p A_q \mid S^1 = A_i A_j, \eta_0) \cdot P(S^1 = A_i A_j \mid \eta_0)}$$

Noting that $P(S^1 = A_i A_j \mid S^2 = A_p A_q, \eta_R) = Rel_R(A_p A_q)_{ij}$, and substituting in the background probabilities:

$$LR = \frac{\sum_{pqij} P(E^1 \mid S^1 = A_i A_j) \cdot P(E^2 \mid S^2 = A_p A_q) \cdot Rel_R(A_p A_q)_{ij} \cdot B_{pq}}{\sum_{pqij} P(E^1 \mid S^1 = A_i A_j) \cdot P(E^2 \mid S^2 = A_p A_q) \cdot B_{pq \mid ij} \cdot B_{ij}}$$

Again by Bayes' theorem we have $P(E^1 \mid S^1 = A_i A_j) = G^1_{ij} P(E)/B_{ij}$, and $P(E^2 \mid S^2 = A_p A_q) = G^2_{pq} P(E)/B_{pq}$ and simplifying with the help of (A3-1) we have

$$LR = \frac{\sum_{ij}\left[\frac{G^1_{ij} \cdot R(\mathbf{G}^2)_{ij}}{B_{ij}}\right]}{\sum_{ij}\left[G^1_{ij} \cdot \sum_{pq} \frac{G^2_{pq} \cdot B_{pq \mid ij}}{B_{pq}}\right]} \qquad (A3\text{-}3)$$

Under the same assumption as Appendix A1 (i.e. HWE or an unconditional population model) we have

$$LR = \sum_{ij} \frac{G^1_{ij} \cdot R(\mathbf{G}^2)_{ij}}{B_{ij}}$$

which establishes theorem 2. In the general case $B_{pq \mid ij} \neq B_{pq}$ (A3-3) should be used.

### A3.1 Example

As an example in the calculation of a LR for a familial relationship, consider a locus with alleles with STR repeat numbers (8, 9, 10, 11) with background frequencies (0.1, 0.2, 0.3, 0.4) respectively. Alice has profile (8,F) and Bob has profile (8, 9). This corresponds to the following matrices (written for convenience in upper-triangular form).

$$\text{Alice} = \begin{bmatrix} 0.1 & 0.2 & 0.3 & 0.4 \\ & 0 & 0 & 0 \\ & & 0 & 0 \\ & & & 0 \end{bmatrix}; \quad \text{Bob} = \begin{bmatrix} 0 & 1 & 0 & 0 \\ & 0 & 0 & 0 \\ & & 0 & 0 \\ & & & 0 \end{bmatrix}; \quad \text{Background} = \begin{bmatrix} 0.01 & 0.04 & 0.06 & 0.08 \\ & 0.04 & 0.12 & 0.16 \\ & & 0.09 & 0.24 \\ & & & 0.16 \end{bmatrix};$$

Suppose we wish to find the LR for the relationship "Alice is a sibling of Bob" (vs "Alice and Bob are unrelated") We first calculate the matrix for "Sibling of Bob" from Equation A2-6

$$Sib(\mathbf{Bob})_{ij} = \sum_{pq} Bob_{pq} \frac{1}{2}(b_i + \delta_{pi}) \cdot \frac{1}{2}(b_j + \delta_{qj})$$

Since the only element of $Bob_{pq} \neq 0$ is $Bob_{10} = 1$ we find

$$Sib(\mathbf{Bob})_{ij} = \frac{1}{4}(b_i + \delta_{0i}) \cdot (b_j + \delta_{1j}) \text{ and therefore}$$

$$Sib(\mathbf{Bob})_{00} = \frac{1}{4}(b_0 + \delta_{00}) \cdot (b_0 + \delta_{10}) = \frac{1}{4}(0.1 + 1) \cdot (0.1 + 0) = 0.0275$$

$$Sib(\mathbf{Bob})_{01} = \frac{1}{4}(b_0 + \delta_{00}) \cdot (b_1 + \delta_{11}) = \frac{1}{4}(0.1 + 1) \cdot (0.2 + 1) = 0.3300 \text{ etc,}$$

giving the components of the matrix shown below.

$$Sib(\mathbf{Bob}) = \begin{bmatrix} 0.0275 & 0.3300 & 0.0825 & 0.1100 \\ 0.0050 & 0.0600 & 0.0150 & 0.0200 \\ 0.0075 & 0.0900 & 0.0225 & 0.0300 \\ 0.0100 & 0.1200 & 0.0300 & 0.0400 \end{bmatrix} \text{ or } \begin{bmatrix} 0.0275 & 0.3350 & 0.0900 & 0.1200 \\ & 0.0600 & 0.1050 & 0.1400 \\ & & 0.0225 & 0.0600 \\ & & & 0.0400 \end{bmatrix}$$

The LR for a sibling relationship between Alice and Bob is obtained by matching "Alice" vs "Sibling of Bob" using Equation (5):

$$\sum \frac{\begin{bmatrix} 0.1 & 0.2 & 0.3 & 0.4 \\ & 0 & 0 & 0 \\ & & 0 & 0 \\ & & & 0 \end{bmatrix} \times \begin{bmatrix} 0.0275 & 0.3350 & 0.0900 & 0.1200 \\ & 0.0600 & 0.1050 & 0.1400 \\ & & 0.0225 & 0.0600 \\ & & & 0.0400 \end{bmatrix}}{\begin{bmatrix} 0.01 & 0.04 & 0.06 & 0.08 \\ & 0.04 & 0.12 & 0.16 \\ & & 0.09 & 0.24 \\ & & & 0.16 \end{bmatrix}}$$

The result of which is

$$\frac{0.1 \times 0.0275}{0.01} + \frac{0.2 \times 0.3350}{0.04} + \frac{0.3 \times 0.09}{0.06} + \frac{0.4 \times 0.12}{0.08} \;=\; 0.275 + 1.675 + 0.45 + 0.6 \;=\; 3$$

# 9  A4 GPMDNA Computer Program

*The first two authors* have developed a computer program GPMDNA to calculate LRs using the method described in this paper. GPMDNA allows the expert to encode an epg as a single GPM for a major contributor, or as two marginal GPMs for a major and a minor as discussed in the main text. An expert using GPMDNA may also encode an *N*-contributor GPM as *2N* allele vectors, each normalised to 1 and representing the probability distribution of one allele. The allele vectors are used by GPMDNA to produce marginal GPMs for contributors to the mixture. Encoding GPMs as products of allele vectors was used in Section 4 for propagating an individual's GPM to his/her relatives. Using allele vectors for interpreting an epg is not completely general, because it implies that allele assignments are independent, but we have had considerable experience of this approach and found it be very flexible, permitting good approximations for a wide range of judgements that an expert might wish make.

We introduce a shorthand notation to specify allele vectors. Any probability not explicitly assigned by the expert is assumed by GPMDNA to be a multiple of background probabilities (denoted '@B'), chosen so that each allele vector is normalised to 1. The B values can be supplied by the software so the expert does not need to know them at the time of encoding. We now illustrate the allele vector encoding and notation using the examples in the main text, assuming a locus with STR alleles 11, 12, ... and associated background probabilities $b_{11}, b_{12}, \ldots$.

Example 1 (See Section 7.2.1)

Having decided there is a single contributor the expert designates the two allele vectors **u** and **v**, and the program generates the components as shown in Table A4-1. So, for example, in case (ii) the expert enters the shorthand notation "11" and "11@0.4/12@0.6" and the program populates the cells of the two vectors and subsequently the cells of the GPM.

| Case | Allele vector | STR allele | | | GPMDNA designation |
|---|---|---|---|---|---|
| | | 11 | 12 | 13 | |
| (i) | **u** | 1 | 0 | 0 | 11 |
| | **v** | $b_{11}$ | $b_{12}$ | $b_{13}$ | F |
| (ii) | **u** | 1 | 0 | 0 | 11 |
| | **v** | 0.4 | 0.6 | 0 | 11@0.4/12@0.6 |
| (iii) | **u** | 1 | 0 | 0 | 11 |
| | **v** | $0.2+0.2b_{11}$ | $0.6+0.2b_{12}$ | $0+0.2b_{13}$ | 11@0.22/12@0.64/13@0.14 |

Table A4-1: Allele vector encoding of Example 1. The pair of rows labelled (i), (ii) or (iii) correspond to the row of Table 1 with the same label. The columns show the allele probability assignments explicitly and in GPMDNA notation. In each case, vector **u** indicates that allele 11 is certainly present, while **v** specifies probabilities for the other allele. In (i) 'F' indicates that background probabilities are assigned to all alleles. In (ii) probabilities of 0.4 and 0.6 are assigned to 11 and 12, corresponding to $\Gamma=0.4$ in Example 1.

Example 2 (See section 7.2.2)

Having decided there are two contributors the expert must designate four alleles : **u**, **v**, **w**, **z**. In this case the expert attempts to deconvolve the major and minor, and assigns **u**, **v** to the major and **w**, **z** to the minor.

| Contributor | Allele vector | STR allele | | | | GPMDNA designation |
|---|---|---|---|---|---|---|
| | | 11 | 12 | 13 | 14 | |
| Major | **u** | 1 | 0 | 0 | 0 | 11 |
| Major | **v** | 0 | 1 | 0 | 0 | 12 |
| Minor | **w** | 0 | 0 | 1 | 0 | 13 |
| Minor | **z** (i) | 0.33 | 0.33 | 0.33 | 0 | 11/12/13@0.33 |
| Minor | **z** (ii) | $\frac{b_{11}}{b_{11}+b_{12}+b_{13}}$ | $\frac{b_{12}}{b_{11}+b_{12}+b_{13}}$ | $\frac{b_{13}}{b_{11}+b_{12}+b_{13}}$ | 0 | 11/12/13@B |
| Minor | **z** (iii) | $\frac{b_{11}}{b_{11}+b_{12}}$ | $\frac{b_{12}}{b_{11}+b_{12}}$ | 0 | 0 | 11/12@B |

Table A4-2: GPMDNA encoding of Example 2 is shown in the rows down to and including **z(i)**. Also shown are two alternative codings of vector **z** that take the background probabilities into account for the fourth allele, either allowing for it to be any of alleles 11, 12 and 13 (vector z(ii), coding "11/12/13@B") or only one of 11 or 12 (vector z(iii) coding "11/12@B").

The 4-allele GPM (shown in Table 3) may be constructed automatically as follows. The GPM for the major is formed from the allele vectors **u** and **v** as in Equation (2), and we will write this as

$$[\mathbf{uv}] = (\mathbf{u}^T\mathbf{v} + \mathbf{v}^T\mathbf{u})/2$$

Similarly the minor is $[\mathbf{wz}]$. We will always permute the vectors for each major or minor component of a profile because we can not distinguish the maternal and paternal alleles. When there is uncertainty about which contributor an allele belongs to, we must perform additional permutations (see examples 3, 4). In this case no more permutations are needed and the 4-allele GPM is just the outer product $A = [\mathbf{uv}] \otimes [\mathbf{wz}]$. In components $A_{ijkl} = (u_i v_j + v_i u_j)(w_k z_l + z_k w_l)/4$. The marginals for the major and minor (Tables 4 and 5) can be got from $A$, or directly as $[\mathbf{uv}]$ and $[\mathbf{wz}]$ respectively.

Example 3 (See section 7.2.2)

In example 3 the GPMDNA designation is straightforward, shown in Table A4-3.

| Contributor | Allele vector | STR repeat number | | | | GPMDNA designation |
|---|---|---|---|---|---|---|
| | | 11 | 12 | 13 | 14 | |
| Major | **u** | 1 | 0 | 0 | 0 | 11 |
| Major/ Minor | **v** | 0 | 1 | 0 | 0 | 12 |
| | **w** | 0 | 0 | 1 | 0 | 13 |
| Minor | **z** | 0 | 0 | 0 | 1 | 14 |

*TableA4-3: GPMDNA encoding of Example 3*

Since the expert has flagged alleles **u** and **v** as belonging to either the major or the minor we must form all possible permutations to get the 4-allele GPM :
$([\mathbf{uv}] \otimes [\mathbf{wz}] + [\mathbf{uw}] \otimes [\mathbf{vz}])/2$ Alternatively we can calculate the 2-allele marginal GPMs directly: $([\mathbf{uv}] + [\mathbf{uw}])/2$ and $([\mathbf{wz}] + [\mathbf{vz}])/2$ for the major and minor respectively. These GPMs are shown in Tables 6, 7 and 8.

Example 4 (See section 7.2.2)

In example 4 there is no attempt to assign alleles to major or minor, so we have:

|  | STR repeat number | | | |
|---|---|---|---|---|
| Allele vector | 11 | 12 | 13 | GPMDNA designation |
| **u** | 1 | 0 | 0 | 11 |
| **v** | 0 | 1 | 0 | 12 |
| **w** | 0 | 0 | 1 | 13 |
| **z** | 0.9 | 0.1 | 0 | 11@0.9/12@0.1 |

*Table A4-4: GPMDNA encoding of Example 4.*

Again we can form the 4-allele GMP of Table 9 by permutation if we need it:
$([\mathbf{uv}]\otimes[\mathbf{wz}]+[\mathbf{uw}]\otimes[\mathbf{vz}]+[\mathbf{uz}]\otimes[\mathbf{vw}]+[\mathbf{vw}]\otimes[\mathbf{uz}]+[\mathbf{vz}]\otimes[\mathbf{uw}]+[\mathbf{wz}]\otimes[\mathbf{uv}])/6$ while forming the permuted product $([\mathbf{uv}]+[\mathbf{uw}]+[\mathbf{uz}]+[\mathbf{vw}]+[\mathbf{vz}]+[\mathbf{wz}])/6$ yields the marginal of Table 10 directly.

GPMDNA Software

The computer program GPMDNA can manage large datasets (tested to over 1 million) and conduct "one to one", "one to many" and "many to many" comparisons. It can store and compare both probabilistic and conventional profiles and it presents results for direct and familial relationships simultaneously. The program, particularly when running on computers with GPGPU cards, is fast enough to allow real time, interactive, investigations of DNA evidence.

The software may be downloaded from: https://github.com/GPMSoftware/GPM.git where the open source licence is described.